# Plasma losses from mirror trap, initiated by microwave radiation under electron cyclotron resonance conditions

D.A. Mansfeld, A.V. Vodopyanov, M.D. Tokman, N.D. Kirukhin, D.I. Yasnov

and V.E. Semenov

Institute of Applied Physics of Russian Academy of Sciences,
603950, 46 Ulyanov str., Nizhny Novgorod, Russia

**Abstract**

Plasma of a stationary ECR discharge sustained by gyrotron radiation at a frequency of 24 GHz and a power of up to 600 W in a mirror magnetic trap was studied. At pressures near the gas breakdown threshold, the extinction effect of an electron-cyclotron discharge is observed with increasing radiation power. When the cyclotron resonance cross section approaches the magnetic plug, the power threshold corresponding to the discharge extinction decreases. These experimental results confirm the presence of the effect of particles precipitation from the trap as a result of electron-cyclotron resonance interaction with gyrotron radiation. The experimental scaling of the extinction power dependence on the position of the ECR cross section in a magnetic trap corresponds to the results of the estimates on the basis of the theory of particle precipitation in the quasilinear diffusion regime.

**Introduction**

The precipitation of particles from open (mirror) magnetic traps in the process of cyclotron resonance interaction with electromagnetic waves plays an important role in the dynamics of both the laboratory [1, 2, 3, 4], and space plasma [5, 6]. The physical mechanism of this phenomenon consists in changing of the pitch angles of the trapped particles by resonant electromagnetic field, which stimulates the transition of particles into the loss cone of a mirror magnetic trap. The sources of electromagnetic radiation can be both cyclotron plasma instabilities [5, 6, 7, 8, 9] and the external generators which are used for the gas ionization and plasma heating [1-4, 10, 11].

The basic concept commonly used to describe the precipitation of particles from the trap under the action of radiation is the quasilinear diffusion of particles in momentum space (see e.g. [12, 13]. In the case of the interaction of charged particles with noise radiation, the diffusion character of the corresponding process appears to be natural, however, for monochromatic radiation at cyclotron resonance in an inhomogeneous magnetic field, the diffusion of particles in the momentum space is also possible. The point is that when moving in an inhomogeneous field of magnetic trap, the particles periodically enter the cyclotron resonance zone, which leads to the occurrence of dynamic chaos [14, 15, 16]. Actually, in this case a "cyclotron" version of the stochastic Fermi acceleration takes place [17].

The effect of electrons precipitation from the mirror-machine under electron-cyclotron discharge conditions was previously studied theoretically in [1] and experimentally in [4]. In these studies, a discharge ignited by monochromatic radiation of powerful – at least 100 kW – gyrotrons was investigated. With these radiation power and typical dimensions of laboratory traps of the order of several tens of centimeters, in addition to the transition of trapped particles to the loss cone with subsequent escape from the trap, an inverse process also plays an important role: the transition of low-energy electrons from the loss cone to the capture region. In this case, the approximation of the "filled" loss cone can be used, which is similar to the so-called regime of strong quasilinear diffusion in the field of noise emission [5, 6] or quasi-gasdynamic model of precipitation of particles from the mirror-machine [18].

Simple estimates show that in typical laboratory mirror traps at the power of electron-cyclotron heating in the range of several hundred watts to several kW, the particle found in the loss cone has only a negligible chance to return to the capture region (see part **III.2**). This is the so-called regime of weak diffusion, when the loss cone in the momentum space should be assumed to be "empty" [5,6]. In this sense, this diffusion regime in the momentum space resembles the well-known "classical" Coulomb losses in mirror traps [18]. The above-mentioned levels of heating power – corresponding to the weak diffusion regime – are the most often used in modern ECR ion sources and other applications [19, 20, 21]. However, as far as we know, the effect of electron-cyclotron interaction with monochromatic radiation on the electron losses from the mirror trap in this regime has not previously been studied experimentally.

In this paper the plasma of a stationary ECR discharge sustained by gyrotron radiation at a frequency of 24 GHz and a power of up to 600 W in a mirror magnetic trap was studied. At pressures near the gas breakdown threshold, the extinction effect of an electron-cyclotron discharge is observed with increasing radiation power. When the cyclotron resonance cross section approaches the magnetic plug, the power threshold corresponding to the discharge extinction decreases (section **I** and **II**). These experimental results confirm the presence of the effect of particles precipitation from the trap as a result of quasilinear diffusion. The experimental scaling of the extinction power dependence on the position of the ECR cross section in a magnetic trap corresponds to the results of the estimates on the basis of the theory of particle precipitation in the weak diffusion regime (section **III**). Some elements of the theory of ECR in an inhomogeneous field are given in the **Appendix**.

**I Experimental scheme and diagnostic methods**

Study of characteristics of electron-cyclotron resonance plasma discharge maintained by continuous microwave radiation of the gyrotron in the mirror magnetic trap was carried out in a setup whose scheme is shown in Fig.1. A technological gyrotron generating continuous electromagnetic radiation with a frequency of 24 GHz and power up to 5 kW was used to create and heat the plasma. The CW emission of the gyrotron propagates into the center of the discharge chamber through a coupled microwave input. The design of the microwave plasma coupling provides transmission of more than 90% of the gyrotron power, also protecting the gyrotron from reflected radiation from plasma and plasma fluxes. A copper grid with holes of 3 mm in diameter is installed at the end of the discharge chamber opposite the microwave inlet, forming a microwave cavity. The gas supply (argon, nitrogen) to the discharge chamber is controlled by a precision leak valve installed next to the microwave inlet. The gas pressure in the chamber is measured by a CC-10 wide-range vacuum gauge. The chamber is evacuated by oil-free forevacuum and turbomolecular pumps, providing the limiting residual pressure in the diagnostic chamber of $(1 \div 2) \times 10^{-7}$ Torr and $(1 \div 2) \times 10^{-6}$ Torr in the area next to the microwave inlet. The operating pressure range is $(5 \div 6) \times 10^{-6}$ Torr. To create ECR conditions, the discharge chamber is placed in the magnetic trap which is formed by two water-cooled Bitter-type magnetic coils. The experimentally measured dependence of the magnitude of the magnetic field on the axis of the chamber on the longitudinal coordinate is shown in Fig.2. The maximum magnetic field in each coil is 1.15 T with a current flow of 740 A; the length of the trap is 13 cm, and the mirror ratio is $R = 6$. The gas breakdown and plasma heating are performed under electron-cyclotron resonance conditions at the fundamental harmonic of the gyrofrequency ($B_{res} = 0.85$ T for 24 GHz).

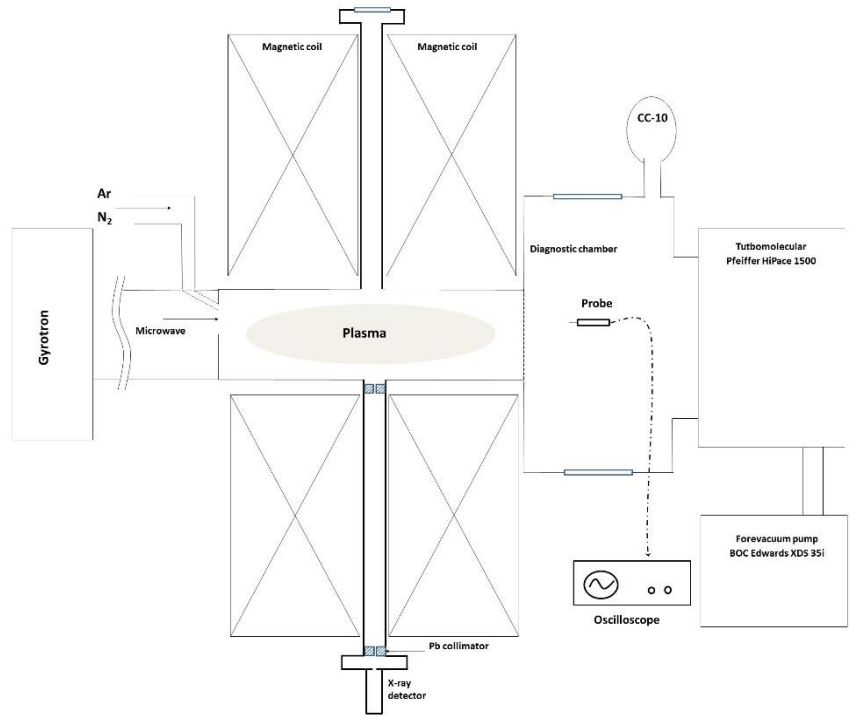

Fig.1. Scheme of the experimental setup

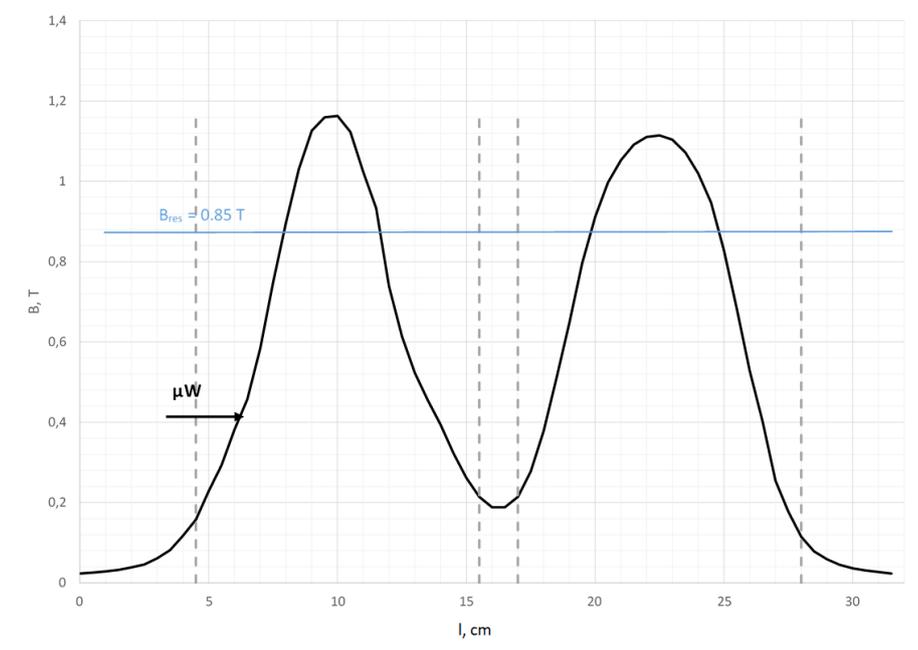

Fig.2. Distribution of the magnetic field along the central axis of the trap. The horizontal line indicates the value of the resonance magnetic field for the frequency of 24 GHz - 0.857 T.

The method for determining plasma density and temperature is based on recording and analyzing the spectrum of the bremsstrahlung radiation emitted by free-free transitions of electrons in the electric field of ions. Assuming a model function of the electron energy distribution, the electron temperature can be determined from the slope of the spectrum on a semilogarithmic scale, and the absolute value of the X-ray intensity can be used to estimate the particle density [22]. To facilitate diagnostics of plasma parameters inside the magnetic trap, two 10 mm diameter flanges with CF40 flanges in the center of the discharge chamber were used. Recording of X-ray quanta in the range 2-15 keV was carried out using a highly sensitive XR-100T detector located 35 cm from the center of the plasma. To reduce the probability of several quanta hitting the detector simultaneously (pileup) and exclude the detection of quanta resulting from electron deceleration on the metal walls of the chamber, the field of view of the detector was limited by lead collimators (1 mm in diameter). As a result, radiation was received from a plasma with a volume of 0.05 cm$^3$.

Measurement of the temperature and plasma density outside the magnetic trap was carried out by a flat Langmuir probe (area of the probe is 0.3 cm$^2$) located in the diagnostic chamber on the axis of the system at a distance of 20 cm from the magnetic mirror. To estimate the plasma density inside the trap, it is assumed that the plasma expansion occurs along the magnetic field lines, and the result was recalculated taking into account the ratio of magnetic flux at the location of the probe to the magnetic flux in the plug.

**II Results of the experiment**

At the maximum magnetic field in the plug (1.15 T at a current of 740 A) and the minimum microwave heating power of 100 W, the gas breakdown occurs only if pressure is more than $5 \times 10^{-5}$ Torr. The appearance of the plasma in the trap is monitored by the electron current to the probe (the probe potential is zero). The features of the plasma formation near the breakdown threshold at a fixed pressure of $7 \times 10^{-5}$ Torr were studied in the experiments.

The discharge is firstly ignited at the minimum power of 50 W and sustained for several minutes. Then the power is gradually increased and at a certain power ($P_{ex}$ – extinction power) the discharge disappears which is monitored by the lack electron current to the probe. With further decrease of power, the discharge is ignited again, starting some power ($P_{ig}$ – ignition power). The effect is well observed in different gases (argon and nitrogen) in the pressure range: $(7 \div 9) \times 10^{-5}$ Torr. The discharge is very unstable with pressures below threshold value, and at pressures above $10^{-4}$ Torr, the discharge is steadily burning and does not extinguish when the power is increased up to 1 kW.

Figure 3 shows a chart of extinction and ignition power vs. the magnetic field of the trap measured at a fixed argon pressure ($7 \times 10^{-55}$ Torr).

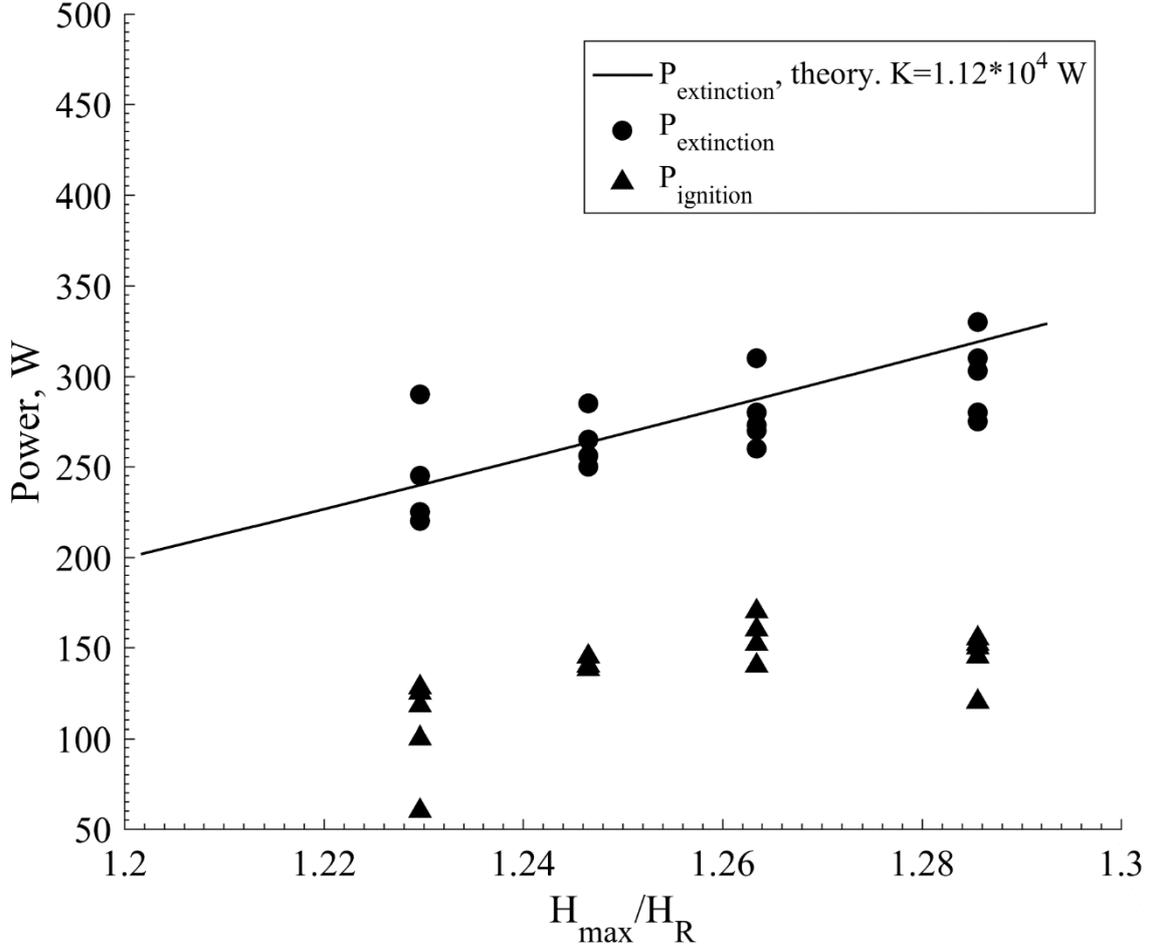

Fig.3. Dependence of the power, which necessary for the extinction of the discharge ($P_{ext}$ – diamonds), and the power needed to resume the discharge ($P_{ig}$ – triangles) on the ratio of magnetic field in the plug to the magnetic field in the resonance cross section. The solid line corresponds to the extinction power calculated for the model considered in section III.

It can be seen from the graph that, with a larger magnetic field, a greater microwave power is required for the discharge to extinguish. At magnetic field values of less than 1.06 T the discharge is very unstable and behave itself very stochastically, and often it is not possible to reignite the discharge by lowering the power down to zero.

With a magnetic field of 1.1 T at the trap mirrors and a fixed gas pressure of $7 \times 10^{-5}$ Torr, the parameters of the plasma emitted from the trap were measured by the Langmuir probe, located in the diagnostic chamber. The investigations were carried out in two different gases (argon and nitrogen) at two characteristic values of the heating power: 120 W (ignition threshold) and 260 W (close to the extinction threshold of the discharge). From the results of processing current-voltage characteristics of the Langmuir probe, shown in Table 1, it is seen that in both the argon plasma and the nitrogen plasma, the electron temperature and the ion saturation current measured by the probe decrease with increasing microwave power. The calculated density of electrons at the mirrors of the trap also decreases more than twofold with increasing power.

| Gas | Power, W | $I_{ionsat}$, nA | $N_{e\_cold}$, 1E9 cm$^{-3}$ | $T_{e\_cold}$, eV |
|---|---|---|---|---|
| Ar | 120 | 3.2 | 2.9 | 5.6 |

|                | 260 | 1.2  | 1.27 | 4.4 |
|----------------|-----|------|------|-----|
| $N_2$          | 120 | 3.3  | 1.45 | 5.8 |
|                | 260 | 0.76 | 0.52 | 3.5 |

Table 1. The results of processing the current-voltage characteristics of the Langmuir probe measured in nitrogen and argon plasma at two power values: the ion saturation current at the probe location, the electron density at the magnetic mirrors, and the electron temperature.

To obtain information on the parameters of the hot electron fraction under the same conditions, the spectra of the X-ray bremsstrahlung radiation of argon and nitrogen plasmas were measured inside the trap. The design of discharge vacuum chamber and the measurement procedure excludes the registration of quanta emitted from the walls. Figures 4 and 5 show graphs of spectral energy density of bremsstrahlung radiation energy vs. photon energy measured in nitrogen and argon plasmas at a heating power of 120 W. In Fig. 5, it should be noted that there is a pronounced maximum at ~ 3 keV, which indicates the excitation of the characteristic spectral lines Kα of argon (2.9 and 3.1 keV).

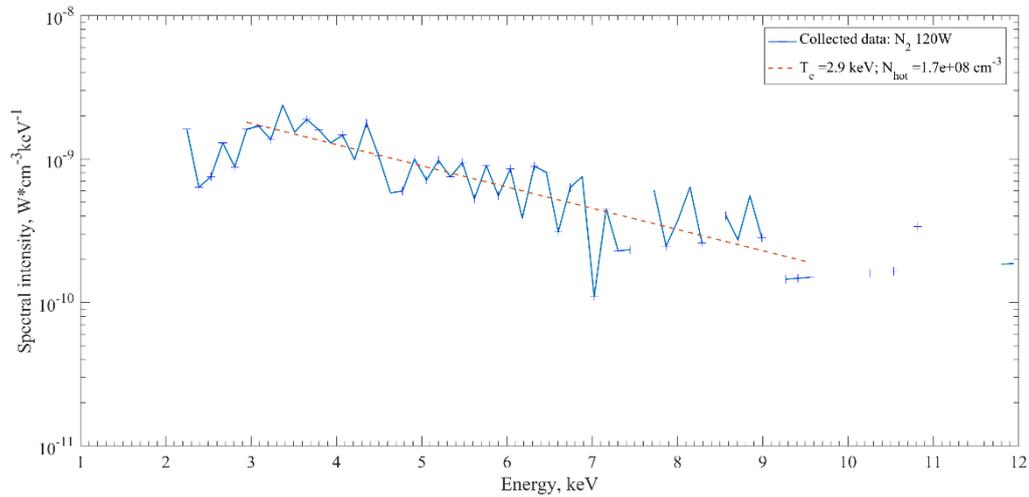

Fig.4. The spectral intensity of plasma X-rays in nitrogen plasma at a microwave power of 120 W. The dashed line is an approximation to the Maxwellian energy distribution function with a temperature of 2.9 keV

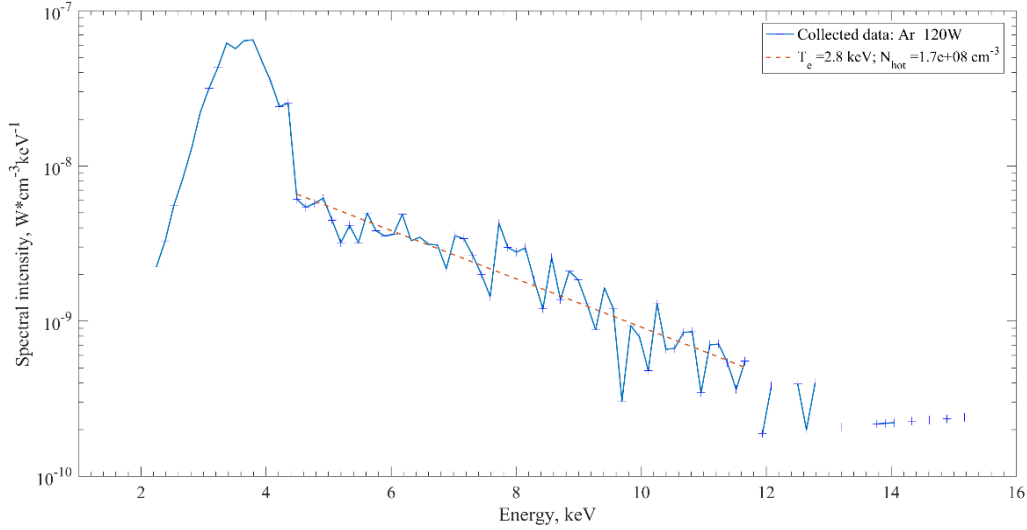

Fig.5. The spectral intensity of plasma X-rays in an argon plasma at a microwave power of 120 W. The dashed line is an approximation to the Maxwellian energy distribution function with a temperature of 2.8 keV

To calculate the plasma parameters under the assumption of the Maxwellian electron energy distribution function, the following formula for the spectral energy density of breaking radiation of electrons in ion field was used [22]:

$$\frac{dE_{ff}}{dv} = C N_e N_i Z_i^2 \left(\frac{\chi_H}{kT_e}\right)^{1/2} e^{-h\nu/kT_e}, \qquad (1)$$

where $E_{ff}$ is the bremsstrahlung radiation energy, $v$ is the bremsstrahlung radiation frequency, $C = 1.7 \cdot 10^{-40}$ erg * cm$^{-3}$ is a constant, $N_e$, $N_i$ are the electron and ion densities, respectively, $Z_i$ is the charge number, $\chi_H$ is the ionization energy of a hydrogen atom, $k$ – is the Boltzmann constant, $T_e$ is the electron temperature, and $h$ is Planck's constant.

Assuming a Maxwellian electron energy distribution, the electron temperature $T_{e\_hot}$ is determined from the slope of the spectrum on a semi-logarithmic scale in the energy region $h\nu \gg T_{e\_hot}$. The plasma density is expressed in terms of the absolute value of the luminescence intensity, taking into account the volume of the radiating region of the plasma falling in the field of view of the detectors. The estimate for the density of hot electrons obtained at the assumption of electron scattering by ions was about $10^{14}$ cm$^{-3}$, which is not realistic, since it exceeds the cutoff density for the heating frequency of 24 GHz. Since the registration of quanta is carried out only from the plasma volume (there is no effect of the chamber walls), we assume that the main contribution to the radiation spectrum is provided by electron deceleration processes on the atomic nuclei. For a rough estimate of the density by deceleration of electrons on nuclei, formula (1) was used with a charge of the nucleus $Z = 18$ for argon atoms and $Z = 7$ for nitrogen atoms. In a weakly ionized plasma, the density of neutral atoms was estimated as $N_a = 3.3 * 10^{16}$ P [Torr] $= 2.3 * 10^{12}$ cm$^{-3}$.

The parameters of hot plasma obtained in the processing of the spectra of bremsstrahlung X-rays are given in Table 2. Since the error in determining the temperature is 10-15%, it can be stated that the temperature of hot electrons in the plasma of both gases does not change with increasing heating power and is 2.9 keV. It can also be argued that, given the error in determining of the density of at least 50%, the density of hot particles practically does not change with increasing power.

| Gas | Power, W | $T_{e\_hot}$, keV | $N_{e\_hot}$, 1E8 cm$^{-3}$ |
|---|---|---|---|
| Ar | 120 | 2.8 | 1.7 |
| | 260 | 3.1 | 1.5 |
| N$_2$ | 120 | 2.9 | 1.7 |
| | 260 | 2.9 | 2.5 |

Table 2. Results of processing of bremsstrahlung spectra measured in nitrogen and argon plasma at two power values: temperature and density of a hot electron component.

According to the probe measurements the plasma density is much less than cut-off value for the heating frequency of 24 GHz, so the effect of discharge extinction with increasing heating power cannot be explained as a result of reflection of microwave emission from plasma. The phenomenon observed in the experiment is naturally associated with the loss of particles from the trap as a result of their interaction with the heating-wave taking into account the fact that ECR heating zone is in proximity of the trap plug. The experimental evidence in Fig.3 shows that a decrease of magnetic field, which means shifting ECR heating region towards the magnetic mirror , results in the lower values of microwave power necessary for discharge extinction. A qualitative theory of the observed effect is given below.

**III Discussion**

**III.1. Quasi-linear equation for the distribution function in the case of monochromatic resonance heating**

Let us consider the motion of charged particles in an inhomogeneous magnetic field $\boldsymbol{H}(\boldsymbol{r})$ and wave field:

$$\boldsymbol{E} = \mathrm{Re}\widetilde{\boldsymbol{E}}(\boldsymbol{r})\mathrm{e}^{i(\int \boldsymbol{k} d\boldsymbol{r} - \omega t)}, \qquad (1)$$

where $\boldsymbol{k}(\boldsymbol{r})$ and $\widetilde{\boldsymbol{E}}(\boldsymbol{r})$ – are the slowly varying in space wave vector and complex amplitude. We believe that the scale of the spatial inhomogeneity of the quantities introduced above is much greater than the wavelength $\frac{2\pi}{k}$ and gyroradius $r_H = \frac{p_\perp}{m\omega_H}$; here $p_\perp$ is the component of the particle momentum $\boldsymbol{p}$ transverse with respect to the magnetic field, $\omega_H = \frac{eH}{mc}$ – nonrelativistic gyrofrequency, $m$ and $e$ – mass and charge of the electron, $c$ – speed of light in the vacuum. Let the Doppler condition be satisfied in some section of the magnetic flux tube [12,13]:

$$\omega - k_\parallel \frac{p_\parallel}{m\gamma} = \frac{N\omega_H}{\gamma}, \qquad (2)$$

where $\omega$ – wave frequency, $N$ – is the number of cyclotron harmonic, $\gamma = \sqrt{1 + \frac{p^2}{m^2c^2}}$ – relativistic factor, $k_\parallel$ and $p_\parallel$ – are the longitudinal components of wave vector and particle momentum with respect to the magnetic field $\boldsymbol{H}$. Neglecting the drift of particles across the magnetic field lines, it is enough to study the dynamics of particles in the variables $p_\perp, \theta; p_\parallel, s$. Here $\theta$ – phase of cyclotron motion[1], $s$ – is the length along the field lines.

Assuming that the distribution over the cyclotron interaction phases $\vartheta = N\theta - \omega t$ becomes random during the oscillations between the magnetic mirrors along the magnetic field, one can arrive at an equation of the Fokker-Planck type describing diffusion in momentum space [14-16]. The equation for the distribution function $f(p_\perp, p_\parallel)$ can be obtained for the fixed cross section of the magnetic tube (see [16]). Operator of quasilinear diffusion in variables $p_{\parallel,\perp}$ is as follows [16]:

$$\left(\frac{\partial f}{\partial t}\right)_{Ql} = \frac{\gamma}{\tau_B(p_\perp, p_\parallel)} \sum_{i,j=\parallel,\perp} \frac{1}{p_i} \frac{\partial}{\partial p_i}\left(D_{ij}(p_{\parallel,\perp}) \frac{1}{p_j} \frac{\partial f}{\partial p_j}\right), \quad (3)$$

where $\tau_B(p_\perp, p_\parallel) = m\gamma \oint \frac{ds}{p_\parallel(s)}$ – period of oscillations between the magnetic mirrors (bounce period), $\frac{p_\parallel(s)}{m\gamma} = \frac{ds}{dt}$. Bounce period is expressed through the values of the momentum components in the selected section. The Eq.(3) is valid for the typical normalization of the distribution function in the kinetic theory:

$$\int_{-\infty}^{\infty} dp_\parallel \int_0^{\infty} p_\perp dp_\perp f(p_\perp, p_\parallel) = n(s), \quad (4)$$

where $n(s)$ – is the particle density at a given point in the coordinate space.

---

[1] Phase angle $\theta_\perp$ is defined by: $p_x = p_\perp \cos\theta$, $p_y = p_\perp \sin\theta$.

The relation between tensor components $D_{\parallel\perp}$ determines the direction of the diffusion flux in the $p_\perp, p_\parallel$ space for the given value $\omega_H(s)$. We note that in the general case the diffusion operator Eq.(3) does not preserve the density $n(s)$, even in the absence of a "loss-cone" in the phase space. The point is that the change in the momenta in the process of stochastic cyclotron acceleration in an inhomogeneous magnetic field leads in general to a redistribution of the particles density along the field lines (see more in [16]).

The direction of the diffusion flux in the momentum space is easiest to establish using the method of quantum analogies (the classical derivation is given in [15,16, 23]). Within the framework of the quantum description, the transverse action in a magnetic field $I_\perp = \frac{p_\perp^2}{2m\omega_H}$ is determined by the Landau levels [24]: $I_\perp = \left(n + \frac{1}{2}\right)\hbar$. Let's consider transitions between levels with numbers $n_{1,2}$. The change in the action upon absorption or emission of a photon is given by: $\Delta I_\perp = \pm(n_2 - n_1)\hbar$. On the other hand, the change in the electron energy in this case is related to the energy of the photon by the relation: $mc^2 \Delta\gamma = \pm\hbar\omega$. As a result, we get: $\omega \Delta I_\perp - Nmc^2 \Delta\gamma = 0$, where $N = n_2 - n_1$ – is the number of cyclotron harmonic, included in the Doppler condition Eq.(2). As a result, we get:

$$\omega I_\perp - Nmc^2 \gamma = \text{const.} \quad (5)$$

We transform Eq. (5), using the connection between the action $I_\perp$ and the transverse component of the momentum $p_\perp$:

$$\frac{\omega}{N\omega_H(s)} \frac{p_\perp^2}{m} - 2mc^2 \gamma = \text{const.}$$

The last equation together with the definition of $\gamma = \sqrt{1 + \frac{p_\perp^2 + p_\parallel^2}{m^2 c^2}}$ determines the relationship between the changes in the momentum components $p_\perp$ and $p_\parallel$ in a given section of the magnetic tube. In the framework of a weakly relativistic approximation (that is, neglecting terms of order $\frac{p^4}{m^4 c^4}$) we get:

$$p_\perp^2 \left(\frac{\omega}{N\omega_H(s)} - 1\right) - p_\parallel^2 = \text{const.}$$

Most simply, this connection is seen in the cross section of the "cold" cyclotron resonance, in which $N\omega_H(s_R) = \omega$:

$$p_\parallel^2 = \text{const.} \quad (6)$$

For the diffusion lines determined by the condition Eq.(6), the general expression Eq.(3) takes the following form in the framework of the weakly relativistic approximation:

$$\left(\frac{\partial f}{\partial t}\right)_{Ql} = \frac{1}{\tau_B(p_\perp,p_\parallel)}\frac{1}{p_\perp}\frac{\partial}{\partial p_\perp}\left(p_\perp \mathfrak{D}(p_\perp,p_\parallel)\frac{\partial f}{\partial p_\perp}\right), \tag{7}$$

where $p_\perp^2 \mathfrak{D} = D_{\perp\perp}$. Here the quantity $\mathfrak{D}(p_\perp, p_\parallel)$ has a simple physical meaning: this is the change of the value of $p_\perp^2$ as a result of the passage through the resonance zone, averaged over the cyclotron interaction phases[2] $\vartheta = N\theta - \omega t$ [1,14-16]. Obviously, for particles in which the region of finite motion does not include the resonance cross section of a magnetic tube, we have $\mathfrak{D} \to 0$. The diffusion lines at the cold resonance cross section and in the bottom of the magnetic well are schematically shown in Figs.6,7.

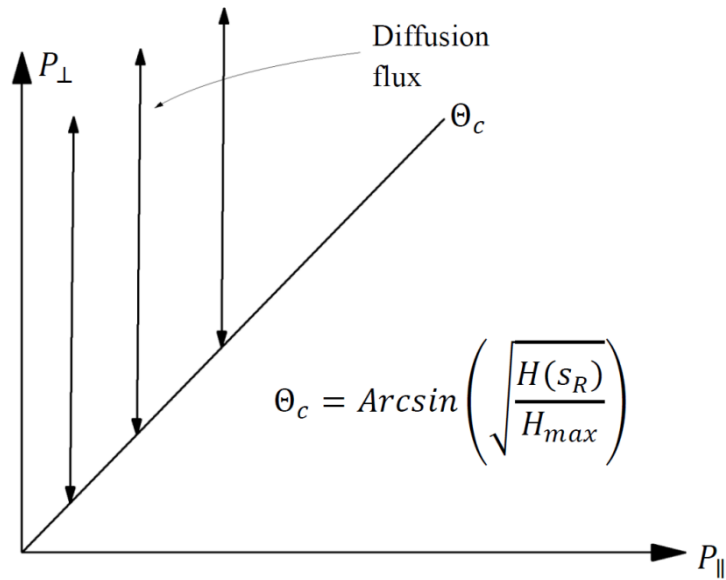

Figure 6. Particle diffusion at the cold resonance cross section

---

[2] Strictly speaking, it is necessary to take into account all the passages of the resonance zones in the vicinity of both magnetic mirrors when moving along a closed trajectory during the period of bounce oscillations.

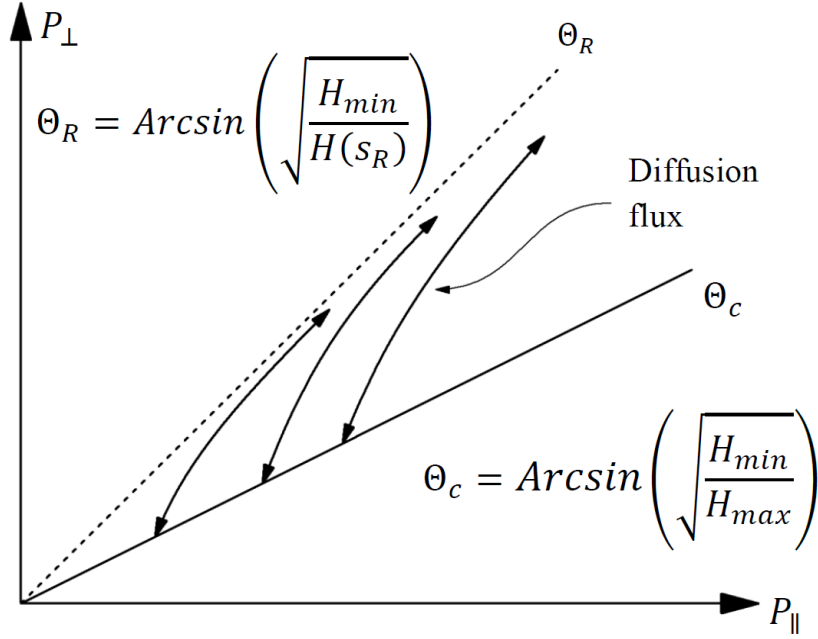

Figure 7. Diffusion lines at the bottom of the magnetic well

The value of $\mathcal{D}$ in the case of resonance at the first gyrofrequency harmonic ($N = 1$) is determined by the expression [25,26]:

$$\mathcal{D} = \frac{1}{2}\left|e\tilde{E}_{(+)} t_{eff}\right|^2, \qquad (8)$$

where $\tilde{E}_{(+)} = \frac{\tilde{E}_x - i\tilde{E}_y}{2}$ – the amplitude of the field component which rotates in the direction of cyclotron rotation of electrons (here the z-axis is oriented along a constant magnetic field). We shall use here the simplest expression for the $t_{eff}$, which corresponds to the span of the resonance zone with a speed $v_\parallel = p_\parallel/m$ [25,26]:

$$t_{eff} = 2\omega^{-1}\sqrt{\pi\omega m l/|p_\parallel|}. \qquad (9)$$

The effective time of cyclotron acceleration $t_{eff}$ in an inhomogeneous magnetic field depends on the scale of the magnetic field inhomogeneity in the resonance cross section $l = \left|\frac{H(s)}{H'_s}\right|_{s=s_R}$. This simple model follows from the general expression given in **Appendix**.

### III.2. General properties of quasilinear diffusion in a mirror trap

The conservation law Eq.(5) determines, in particular, the change in the "turning point" during the cyclotron interaction with the field. In the range of the parameters $\left|\frac{k_\parallel c}{\omega}\right| \geq 1$, which corresponds to the input of radiation along a magnetic, the cross section in which the Doppler condition Eq.(2) is satisfied, is always obtained before the passing the turning point during the finite motion of the particle along the field lines [2,3,23]. However, as the kinetic energy of the particle increases during the stochastic acceleration to relativistic energies, in the case of $\left|\frac{k_\parallel c}{\omega}\right| \geq 1$, the turning point can be displaced in the direction of the magnetic plug up to the escape of the particle from the trap [2, 3]. The energy of the particles, which are leaving the trap is then determined by the relation: $\gamma - 1 \approx 2\left(\frac{H_{max}}{H(s_R)} - 1\right)$. For the parameters of the experiment discussed here, this effect can only determine the characteristic energy of a small high-energy electron fraction (~ 200÷300 keV), which obviously has a little effect on the particle balance in the trap.

All the given above diffusion type equations are obtained in the approximation that the phase of the cyclotron interaction $\vartheta = N\theta - \omega t$ is stochastized during bounce oscillations. Just as for the traditional Fermi acceleration, in the absence of an incidental random effect, such a regime is only possible for not too small amplitudes of the rf field. In the case of cyclotron resonance in a magnetic trap, the following condition is necessary for this [14, 25]:

$$\left|eE_{(+)}\right|t_{eff}\frac{\partial}{\partial p_\perp}\left[\oint \frac{N\omega_H(s)-\omega}{p_\parallel(s)/\gamma m}ds\right] \geq 2\pi. \tag{10}$$

Eq.(10) defines a certain limiting value of the transverse energy, which depends on the amplitude of the high-frequency field, above which the diffusion coefficient tends to zero – the so-called effect of superadiabaticity. The presence of collisions, drift in a system with disturbed axial symmetry, generator frequency instability, etc. will not allow, of course, to realize the effect of superadiabaticity completely. However, one can assume that the coefficient of quasilinear diffusion at the boundary of superadiabaticity at least significantly decreases, which leads to the limitation of the characteristic energy of the weakly relativistic electron fraction (see [1]). On the other hand, the effects mentioned above, which destroy superadiabaticity, may in themselves be the factors limiting the energy of stochastically accelerated particles. This circumstance is especially important in the case of introducing resonant radiation along the magnetic field. As noted above, under the condition $\left|\frac{k_\parallel c}{\omega}\right| \geq 1$ there are no "kinematic" constraints connected with the escape of the resonant cross section from the region of finite motion[3], therefore in this case it is necessary in principle to consider other reasons limiting energy of weakly relativistic hot electrons.

In the framework of this paper we restrict ourselves to the use of experimental values of the characteristic energies of hot electrons. This estimate is justified, since in the range of parameters in which we recorded the experimental dependence of the extinction power on the position of the ECR cross section in a magnetic trap, the characteristic energy of the particles changed insignificantly.

**III.3. Loss of particles stimulated by the resonant field**

The presence of a "loss cone" in momentum space leads to the escape of a part of the electrons with relatively small transverse momenta from the trap. In particular, for the resonance cross section of a magnetic tube, where $\omega = \omega_H(s_R)$, such losses occur for $p_\perp < p_{\perp c} = \frac{p_\parallel}{\sqrt{\frac{H_{max}}{H(s_R)}-1}}$. In this case, the process of quasilinear diffusion leads not only to an increase in the average energy of the particles, but also to their "ejection" from the trap (see Figs.6,7). The experimental situation considered in this paper obviously corresponds to the range of parameters: $p_{\perp c}^2 \gg \mathfrak{D}$. The latter condition ensures that most of the particles in the loss-cone leave the trap during the time of flight between the plugs. In this case it is necessary to solve the diffusion equation $\frac{\partial f}{\partial t} = \left(\frac{\partial f}{\partial t}\right)_{Ql}$ with the standard boundary condition for the "empty" loss cone:

$$f(p_\perp = p_{\perp c}) = 0. \tag{11}$$

---

[3] It is easy to see that for $\left|\frac{k_\parallel c}{\omega}\right| < 1$, i.e. when the radiation is introduced at an angle to the magnetic field (or even at the angle of 90°), there is a corresponding restriction associated with the exit of the resonance cross section given by Eq. (2) from the region of finite motion of the particle or from the region of the wave beam.

To obtain one more boundary condition, we assume the existence of a certain limiting energy $\bar{\epsilon} \approx \sqrt{\bar{p}_\perp^2/2m}$ and require the absence of a particles flux in the momentum space across the boundary $p_\perp = \bar{p}_\perp$:

$$\left(\frac{\partial f}{\partial p_\perp}\right)_{p_\perp = \bar{p}_\perp} = 0. \tag{12}$$

The characteristic loss time $t_{loss}$ can be found by determining the eigenvalue of the diffusion operator:

$$\left(\frac{\partial f}{\partial t}\right)_{Ql} = -\frac{1}{t_{loss}} f \tag{13}$$

for the boundary conditions Eqs (11), (12). In the general case, as a result of such a procedure, we get the value $t_{loss}$, which depends on the conserved quantity $p_\parallel$; in the final expression it makes sense to use some mean (characteristic) value $\bar{p}_\parallel$.

Using Eqs.(7), (13) and the boundary conditions Eqs. (11), (12), we obtain:

$$t_{loss} = \frac{1}{f(\bar{p}_\perp, p_\parallel)} \int_{p_{\perp c}}^{\bar{p}_\perp} \frac{dp_\perp}{p_\perp \mathfrak{D}(p_\perp, p_\parallel)} \int_{p_\perp}^{\bar{p}_\perp} f(p_\perp, p_\parallel) \tau_B(p_\perp, p_\parallel) p_\perp dp_\perp. \tag{14}$$

In the framework of the perturbation method, we substitute in the integral expression Eq. (14) a stationary distribution in the form of the so-called "quasilinear plateau" [1,13]: $f = $ const for $p_{\perp c} < p_\perp \leq \bar{p}_\perp$. Strictly speaking, such method of approximate calculation the eigenvalue Eq.(13) is valid, if the upper boundary of the region of quasilinear diffusion is sufficiently far from the boundary of the loss cone: $\bar{p}_\perp^2 \gg p_{\perp c}^2$. In combination with the condition of the "empty" loss cone, we obtain the following range of applicability of the method used to estimate the characteristic loss time:

$$\bar{p}_\perp^2 \gg \frac{\bar{p}_\parallel^2}{\frac{H_{max}}{H(s_R)} - 1} \gg \mathfrak{D}. \tag{15}$$

As a result, Eq. (14) yields a rather universal expression for the lifetime of the particles precipitated by the resonance rf field from the trap in the framework of the "empty" loss cone model:

$$t_{loss} = \int_{p_{\perp c}}^{\bar{p}_\perp} \frac{dp_\perp}{\mathfrak{D}(p_\perp, p_\parallel) p_\perp} \int_{p_\perp}^{\bar{p}_\perp} \tau_B(p_\perp, p_\parallel) p_\perp dp_\perp. \tag{16}$$

It can be seen that the loss time is determined by integrating the functions $\mathfrak{D}(p_\perp, p_\parallel)$ and $\tau_B(p_\perp, p_\parallel)$ in momentum space, i.e. it depends on the corresponding average values.

For analytical estimates we use the magnetic field profile of the type:

$$H(s) = H_{max} - \frac{H_{max} - H_{min}}{2}\left[1 + \cos\left(\frac{\pi s}{L}\right)\right]; \tag{17}$$

here the magnetic plugs are at the points $s = \pm L$. For the value of $\tau_B$, we use the standard expression for the period of oscillation at the bottom of the magnetic well, given by Eq. (17):

$$\tau_B \approx \frac{4L}{p_\perp/m} \times \frac{1}{\sqrt{\frac{H_{max} - H_{min}}{H(s_R)}}}. \tag{18}$$

Eq. (18) is an approximate expression, while the exact one should tend to ∞ as the turning point tends toward the maximum of the magnetic field. However, the corresponding divergence is only logarithmic[4] and may not be taken into account. Using the expression Eq.(18) for the bounce period, from Eq.(16) we obtain:

$$t_{loss} = \frac{4Lm}{\sqrt{\frac{H_{max} - H_{min}}{H(s_R)}}} \int_{p_{\perp c}}^{\bar{p}_\perp} \frac{\frac{\bar{p}_\perp}{p_\perp} - 1}{\mathfrak{D}(p_\perp, p_\parallel)} dp_\perp. \tag{19}$$

Defining the quasilinear diffusion coefficient $\mathfrak{D}$ with the help of Eqs.(8) and (9), we define the effective time of cyclotron acceleration $t_{eff}(p_\parallel)$ in an inhomogeneous magnetic field for the magnetic field profile of the type Eq.(17) and at resonance near the magnetic plug. Following the expression for the parameter $l$ of the inhomogeneity scale of the magnetic field in the resonance cross section we obtain:

$$l = \left|\frac{H(s)}{H'_s}\right|_{s=s_R} \approx \frac{L}{\pi}\left(1 - \frac{H_{min}}{H_{max}}\right)^{-\frac{1}{2}}\left(1 - \frac{H(s_R)}{H_{max}}\right)^{-\frac{1}{2}}. \tag{20}$$

---
[4] As for a standard nonlinear pendulum.

As a result, using the Eqs.(19),(8),(9) and (20) we get:

$$t_{loss} \approx \ln\left(\frac{\bar{p}_\perp}{p_{\perp c}}\right) \tau_B(\bar{p}_\perp) \frac{\bar{p}_\perp^2}{\mathfrak{D}(\bar{p}_\parallel)}, \tag{21}$$

where $\mathfrak{D}(\bar{p}_\parallel) = \frac{1}{2}\left|eE_{(+)}t_{eff}(\bar{p}_\parallel)\right|^2$; for estimation we should take some characteristic value $\bar{p}_\parallel$.

Thus, in the regime with an "empty" loss cone, the relationship between the lifetime and the characteristic time of the stochastic acceleration $\tau_B \frac{\bar{p}_\perp^2}{\mathfrak{D}}$ is determined by the logarithmic factor.

In Eq.(14) for the lifetime, the only significant factor that depends on the position of the resonance cross section relative to the plug is the gradient of the magnetic field in the resonance cross section. Omitting an insignificant dependence on the mirror ratio in the logarithmic factor we obtain the following scaling law of the loss time $t_{loss}$:

$$t_{loss} \approx \frac{2\omega\bar{p}_\parallel \bar{p}_\perp}{e^2|E_{(+)}|^2} \ln\left(\frac{\bar{p}_\perp}{\bar{p}_\parallel}\right) \sqrt{\left(\frac{H(s_R)}{H_{max}}\right)\left(1 - \frac{H(s_R)}{H_{max}}\right)}. \tag{22}$$

It is interesting to compare the result obtained above with that for the opposite limiting case in which $p_{\perp c}^2 \ll \mathfrak{D}$. The specificity of this situation is the high probability of the particles returning from the loss cone to the confinement region as a result of cyclotron acceleration of particles with relatively small transverse momenta. This situation is close to the so-called the regime of strong diffusion in a resonant field with a noise spectrum [5,6], in which the time of pitch-angle diffusion by the value of the loss cone angle is much less than the transit time between the plugs. For monochromatic radiation the corresponding lifetime was determined in [1]:

$$t_{loss} \approx \left(\frac{H_{max}}{H(s_R)} - 1\right) \tau_B(\bar{p}_\perp) \frac{\bar{p}_\perp^2}{\bar{p}_\parallel^2}.$$

It is seen that in the range of parameter $p_{\perp c}^2 \ll \mathfrak{D}$ the lifetime can depend on the field only through the characteristic value $\bar{p}_\perp$. In [1], $\bar{p}_\perp$ was determined along the superadiabatic boundary – in this case the lifetime increases with increasing rf field. The experiment, which confirmed the developed in [1] theory, was carried out in [4]. In contrast to the radiation power of hundreds of Watts considered in our work, in [4] radiation with a power of more than 100 $kW$ was used.

## 3. The discharge extinction curve

Under the condition $\bar{\epsilon} \gg \epsilon_i$ (here $\epsilon_i$ is the ionization potential), secondary electrons are produced with a characteristic energy spread $\sim\epsilon_i$. In this case, some electrons cannot reach the cyclotron resonance zone, and a part is lost into the loss cone already during the first passage between the plugs. It is clear from Fig.7 that for an isotropic distribution of secondary electrons over pitch angles $\Theta$ ($\Theta = \text{Arccos}\left(\frac{p_\parallel}{p}\right)$) the fraction of secondary electrons $\eta$ involved in the process of stochastic acceleration can be determined as follows:

$$\eta = \int_{\Theta_c}^{\Theta_R} \sin\Theta d\Theta = \sqrt{1 - \frac{H_{min}}{H_{max}}} - \sqrt{1 - \frac{H_{min}}{H(s_R)}}, \tag{23}$$

where $\Theta_c$ is the boundary of the loss cone, $\Theta_R$ is the boundary of the region of the particles that reach the cyclotron resonance zone. In the experiments, the dynamic range of the change in the energy of hot electrons is rather small, which makes it possible to consider the ionization frequency $\nu_i$ to be constant. In the framework of such a model, let us set the discharge extinction boundary by the balance equation $t_{loss}^{-1} = \eta v_i$. Using the relations Eqs. (22), (23), we obtain the following expression for the field intensity $|E_{(+)}|^2$:

$$|E_{(+)}|^2 \approx 2e^{-2}\omega v_i \bar{p}_\| \bar{p}_\perp \ln\left(\frac{\bar{p}_\perp}{\bar{p}_\|}\right) \sqrt{\left(\frac{H(s_R)}{H_{max}}\right)\left(1 - \frac{H(s_R)}{H_{max}}\right)} \left(\sqrt{1 - \frac{H_{min}}{H_{max}}} - \sqrt{1 - \frac{H_{min}}{H(s_R)}}\right) \quad (24)$$

In the present experiment the characteristic energy of the hot electrons $\bar{\epsilon}$ was about $2 \div 3$ keV, the ratio $\frac{H_{max}}{H(s_R)} \approx 1{,}2 \div 1{,}3$. Taking as the characteristic kinetic energy of motion along the magnetic field the value $\bar{p}_\|^2/2m \sim \epsilon_i$ it is easy to verify that the important condition Eq. (15), which we used in calculating the loss time $t_{loss}$, is satisfied with a good margin.

Eq. (24) predicts a drop in discharge extinction power as the resonance cross-section approaches the plug. Using the scaling from Eq. (24) for the extinction power:

$$W = K \sqrt{\left(\frac{H(s_R)}{H_{max}}\right)\left(1 - \frac{H(s_R)}{H_{max}}\right)} \left(\sqrt{1 - \frac{H_{min}}{H_{max}}} - \sqrt{1 - \frac{H_{min}}{H(s_R)}}\right). \quad (25)$$

we obtain a good relative agreement with the experimental data - see Fig.3. The coincidence for absolute values corresponds to the value $K = 1.12 \times 10^4 W \equiv 1.12 \times 10^{11}$ esu. In the framework of the model described above, we obtain the following estimate for the corresponding coefficient: $K \approx \frac{cS\omega v_i \bar{p}_\| \bar{p}_\perp}{\pi e^2} \ln\left(\frac{\bar{p}_\perp}{\bar{p}_\|}\right)$, where $S$ is the area of the aperture of a linearly polarized beam. Ionization frequency $v_i$ was calculated using formula: $v_i = k_i n_p$, where $k_i = 1.176 \times 10^{-7}$ cm$^3$/s – argon ionization constant for $T_{e,hot} = 2{,}8$ keV [27], and $n_p = 2{,}31 \times 10^{12}$ cm$^{-3}$ – density of neutral argon at neutral gas pressure of $7 \times 10^{-5}$ Torr. The typical longitudinal momentum $\bar{p}_\|$ was estimated for ionization potential of Ar ($E_{ion} = 15.76$ eV), the transverse $\bar{p}_\perp$ – for the temperature of hot electrons $T_{e,hot} = 2{,}8$ keV. For the estimates we have taken beam area $S = 11.3$ cm$^2$ which corresponds to the transversal dimension of discharge chamber. As a result, we get a value of $K \approx 1.18 \times 10^{10}$ esu $\equiv 1.18 \times 10^3 W$, that is an order of magnitude smaller than the value corresponding to the experiment. The tenfold divergence for absolute experimental and calculated values of extinction power should not confuse for the qualitive theory and may be due to several factors. In our conclusions we did not take into account the influence of the ambipolar potential, which decreases plasma losses. In addition, the diffusion coefficient, which was defined by expressions (8)-(9) is written very approximately, and additional work should be devoted to its calculation. Also the estimate for the magnitude of the rf field in the resonance zone is made very roughly, since it does not take into account the structure of the field in the cavity in the presence of a plasma. The study of microwave coupling of heating wave with plasma will be done in future work.

We note that in the electron energy range indicated above, the ionization frequency is a decreasing function of the energy. This allows us to expect that both an increase in the power and a decrease in the magnetic field gradient in the resonance cross section can be accompanied by a drop in the ionization frequency (see, for example, [1,4]). In principle, this can also lead to the effect of a drop in the extinction power as the resonance cross-section approaches the plug. However, as noted above, the experimental data on the energy of hot electrons allow this effect to be neglected. Thus, the correspondence between the obtained analytical scaling Eq. (25) to the experimental data confirms the assumption of the important role of the particle extraction effect from the trap under the action of the resonant rf field.

**Acknowledgments**

This work was supported by the grant of Russian Science Foundation # 16-12-10343

**Appendix**

The exact solution of the problem of cyclotron acceleration of a particle in an inhomogeneous magnetic field leads to the expression [25]:

$$t_{eff} = \frac{2\pi\Gamma}{\omega}|\text{Ai}(x)|, \tag{A1}$$

where $\text{Ai}(x)$ – is the Airy function, $x = -\frac{\Gamma\delta_m}{\omega}$, $\Gamma = \left(\frac{2\omega ml}{p_\perp}\right)^{\frac{2}{3}}$, $\delta_m$ – frequencies difference $\omega_H(s) - \omega$ at the turning point where $\frac{ds}{dt} = 0$:

$$\delta_m = \omega\left(\frac{p^2}{p_\perp^2} - 1\right) \tag{A2}$$

If the turning point is sufficiently far from the resonance cross section, when $-x \gg 1$, then the Eq.(9) follows from the corresponding asymptotic of the Airy function. A similar expression is obtained if we assume that the particle passes through the resonance zone at a constant speed, moving first towards the plug, and then back. One can see (see also [1]) that for the applicability of Eq. (9) the characteristic values of the momentum components in the resonance cross section must satisfy the inequality:

$$\left(p_\parallel^2/p_\perp^2\right) \times (2\omega ml/p_\perp)^{\frac{2}{3}} \gg 1. \tag{A.3}$$

In the discussed conditions, the last inequality is certainly satisfied on the boundary of the loss cone, when $p_\parallel^2/p_\perp^2 \sim \left(\frac{H_{max}}{H(s_R)} - 1\right) \sim 0.2$, $p_\perp^2/m \sim \epsilon_i$. Thus, the expressions used to determine the coefficient of quasilinear diffusion are certainly correct near the boundary of the loss cone, where, in fact, the flow of particles leaving the trap is formed.